\documentclass[sigconf]{acmart}

\AtBeginDocument{%
}

\copyrightyear{2026}
\acmYear{2026}
\setcopyright{cc}
\setcctype{by}
\acmConference[DIS Companion '26]{Designing Interactive Systems Conference}{June 13--17, 2026}{Singapore, Singapore}
\acmBooktitle{Designing Interactive Systems Conference (DIS Companion '26), June 13--17, 2026, Singapore, Singapore}
\acmDOI{10.1145/3802974.3809475}
\acmISBN{979-8-4007-2632-3/2026/06}

\usepackage{graphicx}
\usepackage{float}
\usepackage{subfigure}
\usepackage{subcaption}
\usepackage{hyperref}

\begin{document}

\title{\textsc{SpatialPrompt}: XR-Based Spatial Intent Expression as Executable Constraints for AI Generative 3D Design}

\author{Yichen Andy Yu}
\orcid{0009-0001-0175-3253}
\affiliation{%
  \institution{Department of Computer Science \\ North Carolina State University}
  \city{Raleigh}
  \state{North Carolina}
  \country{USA}
}
\email{yyu55@ncsu.edu}

\author{Wanru Li}
\orcid{0009-0002-8988-5572}
\affiliation{%
  \institution{CIT Interdisciplinary \\ Carnegie Mellon University}
  \city{Pittsburgh}
  \state{Pennsylvania}
  \country{USA}
}
\email{wanrul@alumni.cmu.edu}

\author{Qiaoran Wang}
%% \authornotemark[1]
\orcid{0009-0006-4483-1107}
\affiliation{%
  \institution{National University of Singapore}
  \city{Singapore}
  \country{Singapore}
}
\email{violetforever999@gmail.com}

\author{Jymon Ross}
\orcid{0009-0006-2726-2651}
\affiliation{%
  \institution{Department of Computer Science \\ North Carolina State University}
  \city{Raleigh}
  \state{North Carolina}
  \country{USA}
}
\email{jross5@ncsu.edu}

\author{Gavin Johnson}
\orcid{0009-0002-9235-4596}
\affiliation{%
  \institution{Department of Computer Science \\ North Carolina State University}
  \city{Raleigh}
  \state{North Carolina}
  \country{USA}
}
\email{grjohns5@ncsu.edu}

\author{Mandy Lui}
\orcid{0009-0005-8913-9942}
\affiliation{%
  \institution{Department of Computer Science \\ University of Rochester}
  \city{Rochester}
  \state{New York}
  \country{USA}
}
\email{mlui@u.rochester.edu}

\author{Qiao Jin}
\orcid{0000-0001-5493-1343}
\affiliation{%
  \institution{Department of Computer Science \\ North Carolina State University}
  \city{Raleigh}
  \state{North Carolina}
  \country{USA}
}
\email{qjin4@ncsu.edu}

\renewcommand{\shortauthors}{Yu et al., with Jin}

\begin{abstract}
We present \textsc{SpatialPrompt}, an Extended Reality(XR) system that turns spatial sketches into executable constraints for controllable 3D generation. Users draw rough structures with a 3D pen and add voice prompts for semantic and stylistic intent. The system supports iterative refinement and synchronous co-creation in shared space with color-coded contributions. Implemented on Apple Vision Pro with Logitech Muse and Meshy, a heuristic evaluation suggests that the workflow is intuitive and supports shared understanding in collaborative creation, while revealing needs for faster generation and clearer feedback.
\end{abstract}

\begin{CCSXML}
<ccs2012>
   <concept>
       <concept_id>10003120.10003121.10003124.10010866</concept_id>
       <concept_desc>Human-centered computing~Virtual reality</concept_desc>
       <concept_significance>500</concept_significance>
       </concept>
   <concept>
       <concept_id>10003120.10003121.10003124.10010392</concept_id>
       <concept_desc>Human-centered computing~Mixed / augmented reality</concept_desc>
       <concept_significance>500</concept_significance>
       </concept>
</ccs2012>
\end{CCSXML}

\ccsdesc[500]{Human-centered computing~Virtual reality}
\ccsdesc[500]{Human-centered computing~Mixed / augmented reality}
\ccsdesc[500]{Human-centered computing~Interaction paradigms}
\ccsdesc[300]{Computing methodologies~Artificial intelligence}

\keywords{Generative AI, Augmented Reality, Spatial Language}

\begin{teaserfigure}
  \includegraphics[width=\textwidth]{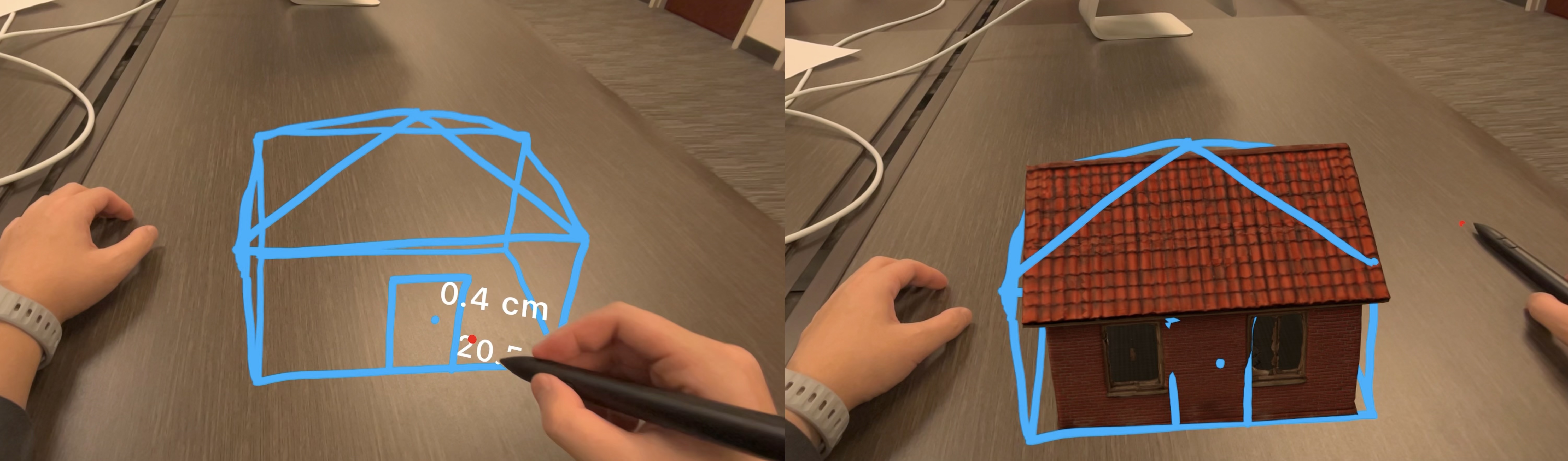}
  \caption{Overview of \textsc{SpatialPrompt}, where users sketch spatial structures with a 3D pen and provide voice prompts in XR to generate and iteratively refine corresponding 3D assets in a tabletop augmented reality workflow.}
  \label{fig:teaser}
\end{teaserfigure}

\maketitle

\section{Introduction}
Recent advances in generative AI have lowered barriers to 3D content creation, enabling users to generate assets from text or image prompts~\cite{poole2022dreamfusion,lin2023magic3d,nichol2022pointe,jun2023shape,jiang2024text3dsurvey}. However, expressing \emph{spatial} design intent remains difficult: non-professional users often struggle to specify structure, proportion, and layout precisely enough for generative systems to execute~\cite{kleinsmann2008sharedunderstanding,buxton2007sketching}. This challenge is amplified in collaborative settings, where misalignment between non-experts and designers can lead to repeated revisions and costly back-and-forth~\cite{kleinsmann2008sharedunderstanding}.

Most generative interfaces rely on language or 2D references, which convey semantics and appearance but often underspecify 3D structure~\cite{jiang2024text3dsurvey,liu2024sketchdream}. We therefore explore an alternative input paradigm based on \emph{spatial intent expression} in Extended Reality (XR). We present \textsc{SpatialPrompt}, an XR system that lets users sketch rough 3D structures with a spatial pen and add voice prompts for semantic and stylistic intent~\cite{keefe2007drawingonair,joundi2020vrsketching,tong2024ms2meshxr}. Rather than treating sketches as final models, \textsc{SpatialPrompt} compiles them into \emph{executable constraints} that guide generation~\cite{sutherland1963sketchpad,zou2022gcsreview}, representing a spatial prompt as lightweight geometric structure and operations that enforce intended boundaries, scale, and organization.

Our goal is not to replace professional designers, but to support communication and alignment in early-stage co-creation by providing a shared 3D reference that is both editable and generative.

\section{Related Work}
\subsection{AI-to-3D Generation}

Generative 3D pipelines have advanced rapidly, but most interfaces still rely on text and other 2D semantic conditions, leaving limited control over 3D structure~\cite{jiang2024text3dsurvey}. We summarize three common input paradigms and their limitations for design communication.
Text prompts can produce high-quality assets~\cite{poole2022dreamfusion,lin2023magic3d}, but natural language often underspecifies scale, proportion, topology, and part relationships~\cite{jiang2024text3dsurvey}. Users therefore depend on iterative prompt tuning, which can amplify ambiguity in collaborative settings. Single-image methods convey appearance well~\cite{liu2023zero123}, but remain limited for communicating intended structure and spatial constraints, especially for creation rather than reconstruction~\cite{jiang2024text3dsurvey}. Video reconstruction pipelines such as NeRF and 3D Gaussian Splatting recover scenes from observed imagery~\cite{mildenhall2020nerf,kerbl2023gaussiansplatting}, but primarily support capturing existing environments rather than specifying new designs.

Overall, these paradigms struggle to represent structural relationships, scale, and spatial organization explicitly~\cite{jiang2024text3dsurvey}. We therefore explore XR-native spatial structure as a prompt modality and compile it into executable constraints to improve controllability and collaborative design communication.

\subsection{Designing Methods in XR}

Prior work explores how XR supports early-stage design through embodied interaction and shared spatial understanding. We summarize three related directions. XR sketching tools let designers externalize ideas in 3D space during early ideation and prototyping~\cite{joundi2020vrsketching,keefe2007drawingonair}. While effective for expressing form and proportion, turning sketches into production-ready assets still typically requires expert modeling or extensive back-and-forth. Studies of VR design workflows analyze iteration strategies and interaction patterns, showing how immersive affordances shape creative processes~\cite{wang2024vrdesignprocess}. However, they do not focus on compiling spatial expressions into controllable constraints for AI-assisted synthesis. Collaborative XR work demonstrates how shared spatial representations support coordination and decision-making~\cite{dorta2019socialvr,bakk2025socialvrcodesign}. Yet few systems integrate collaborative spatial expression with generative AI in a closed-loop pipeline where spatial artifacts act as executable constraints for controllable generation.

\section{\textsc{SpatialPrompt} System}

\subsection{Design Goals}
Based on the research gaps identified in the related work, our design goals are listed as follows:

\begin{itemize}
\item \textbf{Enable Spatial Intent Expression for Non-Experts.}
Lower the barrier of 3D design by allowing users without professional modeling expertise to externalize ideas through embodied spatial interaction. The system should support intuitive expression of structure, proportion, and volume using direct spatial sketching rather than requiring technical modeling operations or specialized terminology.

\item \textbf{Support Controllable Generative Design via Spatial Constraints.}
Transform user-created spatial structures into executable constraints that guide generative models. Instead of relying solely on text or image prompts, the system should leverage spatial input to define boundaries, scale relationships, and structural cues, improving controllability and reducing unpredictability in generated results~\cite{sutherland1963sketchpad,zou2022gcsreview}.

\item \textbf{Facilitate Iterative Human-AI Co-Creation.}
Support a continuous design loop in which users can refine spatial sketches, update constraints, and regenerate models incrementally. The system should enable rapid feedback cycles that allow users to progressively converge toward desired outcomes while maintaining agency and interpretability.

%%\item \textbf{Support Collaborative Co-Creation and Shared Spatial Understanding.}
%Enable multiple participants to collaboratively interact within a shared immersive environment. The system should promote shared spatial cognition by allowing co-editing, clear contribution differentiation, and joint manipulation of spatial prompts, helping align perspectives and reduce miscommunication between collaborators~\cite{gutwin2002workspaceawareness,sereno2022collabar,cidota2016workspacear}.
\end{itemize}

\subsection{System Design Overview}
\begin{figure}[t]
  \centering
  \includegraphics[width=\linewidth]{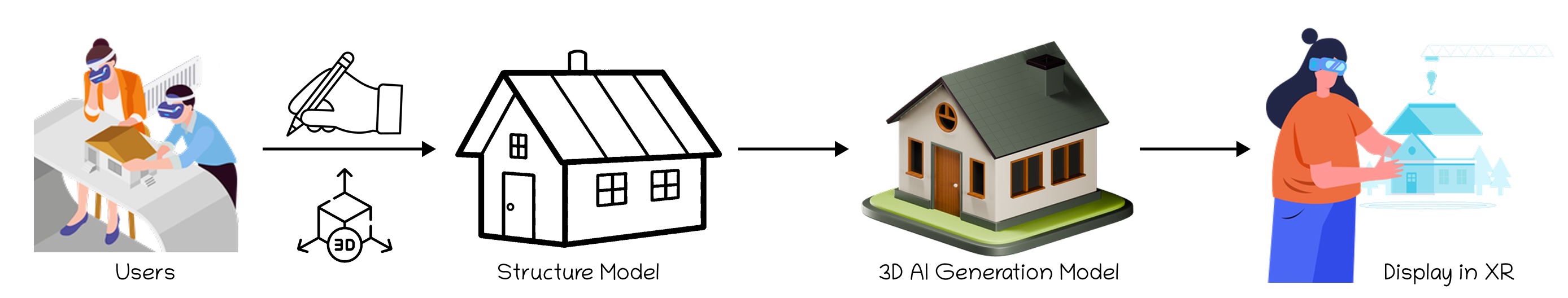}
  \caption{End-to-end workflow of \textsc{SpatialPrompt}: users create a spatial structure model in XR, which is compiled into executable constraints to condition a 3D generation backend, and the generated asset is displayed and refined in XR.}
  \Description{A left-to-right pipeline diagram showing users sketching a structure model, sending it to a 3D AI generation model, and displaying the result in XR.}
  \label{fig:overview}
\end{figure}

Addressing the gap between users’ spatial intent and what generative 3D systems can reliably execute, our system treats XR interaction as a medium for spatial communication (Figure~\ref{fig:overview}). Rather than lowering modeling barriers alone, it helps users externalize spatial concepts, compile them into generative constraints, and iteratively refine designs through collaboration.

\subsubsection{Enable Spatial Intent Expression for Non-Experts}

Users directly sketch rough outlines and structural frameworks of objects within XR using the 3D Pencil, rapidly generating an editable, precise 3D sketch model. Creation occurs at the desktop or any spatial location; the system allows users to leverage the real environment as a scale reference, quickly calibrating model dimensions through relative size relationships with surrounding objects. This enables early structural expression and scale alignment without precise measurements or professional modeling skills. Through these spatial interaction mechanisms, the system supports users in directly conveying structure, proportion, and volume within a shared 3D environment. This empowers non-professionals to efficiently externalize design intent while establishing clear spatial constraints for subsequent generative completion~\cite{keefe2007drawingonair,joundi2020vrsketching}.

Simultaneously, to address the limitations of purely geometric sketches in conveying semantic and design preferences, the system further supports voice prompts as supplementary input. In our interaction design, the 3D Pen handles spatial structural intent while voice prompts convey semantic and stylistic intent. Together, they form composite prompts for generative models, enabling non-professional users to articulate both how it should look and what it should be in everyday language~\cite{tong2024ms2meshxr}. This reduces miscommunication with designers caused by imprecise language or insufficient references~\cite{kleinsmann2008sharedunderstanding}.

\subsubsection{Enabling Controlled Generative Design Through Spatial Constraints}

Existing generative 3D interfaces predominantly rely on text or image prompts, yet such inputs often struggle to precisely convey spatial structural information~\cite{jiang2024text3dsurvey}. This frequently leads to generated outputs deviating from user expectations in form, layout, or structure. To overcome this limitation, our system redefines spatial inputs generated by users in XR, including 3D wireframe sketches, structural frameworks, and imported partial geometries, such as constraints and control signals within the generation process, rather than as final output models~\cite{sutherland1963sketchpad,zou2022gcsreview}.

Specifically, the system interprets spatial inputs as structural guidance. Wireframe sketches define primary contours and critical transitions; structural frameworks indicate large-scale volumetric structures and spatial occupancy; and local geometries serve as anchors for local shapes or functional components. Based on this, the system forms a set of generation-oriented spatial constraints, including but not limited to: bounding regions, relative scale and proportion, structural scaffolds, and spatial organization cues. These constraints collectively determine which parts must be retained and which parts allow AI-driven completion, transforming generative freedom from completely uncontrollable to exploration within defined boundaries.

This design elevates geometric inputs into spatial prompts, enabling users to guide generation outcomes through simple spatial operations. By modifying the spatial structure itself, the system constrains the direction of generation. By anchoring the generation process to explicit spatial constraints, the system significantly reduces unpredictability and trial-and-error costs, enabling more purposeful design exploration while providing a shareable, traceable structural basis for subsequent designer communication~\cite{sutherland1963sketchpad}.

\subsection{Implementation}
The system is implemented using Apple Vision Pro and developed with Xcode. Users can create on desktops or any spatial location; the system anchors drawings to the world coordinate system, ensuring spatial consistency between sketches and models when moving the viewpoint. Logitech Muse serves as the spatial pen input device. The system maps Muse's pointing and pose inputs to a 3D pen, enabling users to directly draw three-dimensional lines on the desktop or in mid-air to express structural outlines and key frameworks. To enhance drawing stability and editability. During the generation phase, we combine the user's drawn 3D model structure with a voice prompt to form a generation request: the sketch provides spatial structure and scale constraints, while the voice prompt supplements semantic meaning, material properties, and stylistic preferences. We utilize the Meshy API as the backend generation service, converting this composite input into usable 3D assets.

\section{Heuristic Evaluation}
We conducted a formative heuristic evaluation with three laboratory members to identify early usability issues and gather feedback for iterative improvements~\cite{nielsen1994heuristic,nielsen1993model}. All participants had at least two years of XR development or design experience and backgrounds in human--computer interaction(HCI) or interaction design. After a brief introduction to the workflow, participants completed two tasks. First is a single-user 10-minute session in which they sketched a target object, issued voice prompts , triggered generation, and performed at least one sketch-edit regeneration cycle and next is a collaborative 10-minute session in which two participants co-edited the same sketch in a shared space using color-coded strokes and jointly triggered regeneration. Participants provided numerical ratings on intent expressiveness, generated model quality, and expectation alignment, and then completed a 15-minute interviews to explain ratings and report issues. Overall, participants found spatial sketching effective for quickly externalizing structure and considered the sketch+voice workflow intuitive for conveying both geometry and semantic intent; however, they noted occasional inconsistencies in details or proportions. In collaboration, color-coded contributions supported attribution and coordination and reduced reliance on verbal explanation~\cite{gutwin2002workspaceawareness,cidota2016workspacear}, while feedback highlighted the need for faster generation and clearer visual feedback during collaborative editing.

\section{Limitations and Future Work}

Although the proposed system demonstrates potential for bridging the communication gap between collaborative creation by integrating spatial interaction with generative AI, several limitations remain, presenting opportunities for future research.

Also, the generation process relies on external AI services, potentially leading to fluctuations in output quality and response times. While spatial constraints enhance controllability, generated models may still deviate from user expectations when semantic prompts are ambiguous or structural sketches are incomplete. Future research could explore more transparent generation pipelines, interactive constraint visualization, or real-time preview mechanisms to clearly demonstrate how user inputs influence outcomes.

Furthermore, as an early heuristic study, this evaluation involved only a small number of participants with VR and HCI experience. While this approach efficiently identifies usability issues, broader research encompassing non-expert users and professional designers is needed to validate the system's effectiveness in real design communication scenarios. Future research may explore long-term collaborative applications, examining how spatial intent expression influences negotiation dynamics and decision-making processes between collaborative creation.

Collectively, these directions highlight the potential for expanding spatial cues into a novel generative design interaction paradigm, enabling it to transcend early conceptual phases and evolve into more robust collaborative design workflows.

\bibliographystyle{ACM-Reference-Format}
\bibliography{reference}

\end{document}